\documentclass[11pt,reqno,a4,onecolumn,twoside,final,spanish]{article}

\usepackage{amsthm,amsmath,amssymb,latexsym,amsfonts,amscd} 
\usepackage{latexsym}
\usepackage{url}
\usepackage[T1]{fontenc}
\usepackage{ae}
\usepackage{aecompl}
\usepackage[small,sc,hang,bf]{caption} 
\usepackage{graphicx,color,ae,fancyvrb}
\usepackage{pst-all}
\usepackage{subfig}
\usepackage{float} 
\usepackage{verbatim}
\usepackage{enumerate}
\usepackage{array}
\usepackage{longtable}
\usepackage{multicol}
\usepackage{cancel}
\newcommand*\chancery{\fontfamily{pzc}\selectfont}

\newlength\mytemplen
\newsavebox\mytempbox

\makeatletter
\newcommand\mybluebox{%
    \@ifnextchar[
       {\@mybluebox}%
       {\@mybluebox[0pt]}}

\def\@mybluebox[#1]{%
    \@ifnextchar[
       {\@@mybluebox[#1]}%
       {\@@mybluebox[#1][0pt]}}

\def\@@mybluebox[#1][#2]#3{
    \sbox\mytempbox{#3}%
    \mytemplen\ht\mytempbox
    \advance\mytemplen #1\relax
    \ht\mytempbox\mytemplen
    \mytemplen\dp\mytempbox
    \advance\mytemplen #2\relax
    \dp\mytempbox\mytemplen
    \colorbox{myblue}{\hspace{1em}\usebox{\mytempbox}\hspace{1em}}}

\makeatother

\usepackage{natbib}

\usepackage{hyperref}

\usepackage{geometry}
\usepackage{calc}
	
\makeatletter
\renewcommand{\section}{\@startsection{section}{1}{0pt}{-3ex plus -1ex minus 0ex}{2ex plus 0ex}{\bf}}
\makeatother

\makeatletter
\renewcommand{\subsection}{\@startsection{subsection}{1}{0pt}{-2ex plus -1ex minus 0ex}{2ex plus 0ex}{\bf}}
\makeatother

\theoremstyle{definition}

\theoremstyle{remark}

\begin{document}

\renewcommand{\tablename}{Tabla}
\renewcommand{\figurename}{Figura}
\noindent

\begin{flushleft}
\textsl {\chancery  Memorias de la Primera Escuela de Astroestad\'istica: M\'etodos Bayesianos en Cosmolog\'ia}\\
\vspace{-0.1cm}{\chancery  9 al 13 Junio de 2014.  Bogot\'a D.C., Colombia }\\
\textsl {\scriptsize Editor: H\'ector J. Hort\'ua}\\
\href{https://www.dropbox.com/sh/nh0nbydi0lp81ha/AACJNr09cXSEFGPeFK4M3v9Pa}{\tiny {\blue Material suplementario}}
\end{flushleft}



\thispagestyle{plain}\def\@roman#1{\romannumeral #1}



\begin{center}\Large\bfseries Determinaci\'on de los par\'ametros cosmol\'ogicos mediante el efecto de lente gravitacional d\'ebil. El caso de $\Omega_m$, $\sigma_8$. \end{center}
\begin{center}\normalsize\bfseries  Cosmological parameter estimation from weak lensing. The case of $\Omega_m$, $\sigma_8$.\end{center}

\begin{center}
\small
\textsc{Leonardo Casta\~neda\footnotemark[1]}
\footnotetext[1]{Observatorio Astron\'omico Nacional, Universidad Nacional de Colombia, Bogot\'a D.C. E-mail: \url{lcastanedac@unal.edu.co}}

\end{center}

\noindent\\[1mm]
{\small
\centerline{\bfseries Resumen}\\
La propagaci\'on de la luz en el universo con estructuras las cuales amplifican  y modi\-fican la forma de las galaxias distantes, produciendo una correlaci\'on entre la densidad de galaxias pr\'oximas y lejanas, es un fen\'omeno que ha ganado importancia en  cosmolog\'ia para la determinaci\'on de par\'ametros cosmol\'ogicos como es el caso del modelo estandar ($\Lambda$CDM). Durante el presente art\'iculo se discutir\'a la estimaci\'on de la funci\'on de correlaci\'on de dos puntos en el cizallamiento gravitacional producido por la estructura a gran escala del universo (usualmente conocido como \emph{cosmic shear} $\gamma$). Se comparar\'an los resultados proporcionados de las lentes gravitacionales con el empleo de otras alternativas como la abundancia de c\'umulos de galaxias. Tambi\'en se describen algunos software utilizados en el estudio de lentes gravitacionales para determinar modelos de distribuci\'on de masa y generaci\'on de im\'agenes. \\
{\footnotesize
\textbf{Palabras clave:}
Lente gravitacional d\'ebil, Estimaci\'on de par\'ametros cosmol\'ogicos.
\\
\noindent\\[1mm]
{\small
\centerline{\bfseries Abstract}\\

Propagation of light in the universe with structure which  amplify and modify the shape of distant galaxies, producing a correlation between nearby and distant density of galaxies, is a phenomena very important in cosmology for determining cosmological parameters as the $\Lambda$CDM.
In this paper, we discuss the estimation of the two point correlation function in the gravitational shear produced by the large scale structure. We will compare the result
given by gravitational lensing with the use of another alternatives such as a counting galaxy clusters. We also describe some software used in the gravitational lensing study  for determining mass distribution models and images formation.   \\

{\footnotesize
}

\newpage
\section{Introducción}
El modelos est\'andar cosmol\'ogico descansa en dos tesis te\'oricas fundamentales,
\begin{itemize}
\item{El universo es en grandes escalas \Big($100\textit{Mpc}$ y mayores \Big) estad\'{\i}sticamente homog\'eneo e isotr\'opico.}
\item{La gravitaci\'on es la interacci\'on fundamental que gobierna la din\'amica a gran escala del universo. }
\end{itemize}
Basados en las consideraciones anteriores, el paradigma actual de la cosmolog\'{\i}a denominado $\Lambda\textit{CDM}$ por sus siglas en ingl\'es ($\Lambda$-Cold-Dark-Matter) es un campo de grandes avances cient\'{\i}ficos y enormes retos a nivel te\'orico y observacional los cuales deben ser direccionados vali\'endose de diversas t\'ecnicas en diferentes ramas de la f\'{\i}sica, matem\'atica y estad\'{\i}stica principalmente.

La cosmolog\'{\i}a es una ciencia observacional principalmente, para lo cual se debe basar en modelos te\'oricos para la interpretaci\'on de los datos obtenidos. La fuente principal de informaci\'on es la radiaci\'on electromagn\'etica y la regi\'on observable del universo, delimitada por la propagaci\'on de luz, se conoce como \emph{cono de luz pasado.} En un universo con estructuras, la propagaci\'on de luz y como consecuencia la determinaci\'on del cono de luz pasado es un problema estad\'{\i}stico por naturaleza.
Por otro lado, el contenido de momento-energ\'{\i}a determinada la din\'amica y la geometr\'{\i}a del universo en una teor\'{\i}a como la Relatividad General de Einstein, la cual es a la fecha la mejor descripci\'on de la gravedad conocida. El modelo del universo determinado a partir de  las ecuaciones de la Relatividad General (o extensiones de la misma), demanda el conocimiento de un conjunto de par\'ametros, denominados \textbf{par\'ametros cosmol\'ogicos} y usualmente denotados en la literatura como $\Big\{\Omega_m,\Omega_{\Lambda},\Omega_{b},H_{0},\Gamma, \sigma_8, q_0, \tau_{\textit{reion}}\Big\}$. Cuando se hace referencia a un modelo cosmol\'ogico se entiende el modelo determinado por un conjunto particular de par\'ametros de los anteriormente mencionados.

Una de las grandes diferencias de la cosmolog\'{\i}a con el resto de ciencias consiste en que en cosmolog\'{\i}a, por definici\'on, \'unicamente poseemos una sola realizaci\'on de nuestro de nuestro universo. Los par\'ametros cosmol\'ogicos deben ser determinados con el uso de esa \'unica realizaci\'on, en otras palabras, no tenemos la posibilidad de repetir los experimentos y determinar el valor de los par\'ametros cosmol\'ogicos mediante este n\'umero de repeticiones. Debido a este hecho, las t\'ecnicas de estad\'{\i}stica Bayesiana ofrecen una oportunidad para la estimaci\'on de los par\'ametros cosmol\'ogicos, \cite{tereno}. Para realizar dicha tarea, durante las d\'ecadas pasadas se han desarrollado grandes pruebas observacionales como lo constituyen los cat\'alogos de galaxias, entre los mas destacados est\'an el Sloan-Digital-Sky-Survey y el $2d-$Galaxy Redshift Survey. Con dichos cat\'alogos se ha estimado con gran precisi\'on la funci\'on de correlaci\'on gal\'actica. Adem\'as de dichos esfuerzos por tener un mapa completo de la distribuci\'on de galaxias, otras pruebas como la utilizaci\'on de candelas est\'andar denominadas  supernovas (SNIa) y las lentes gravitacionales han proporcionado t\'ecnicas para la determinaci\'on precisa de distancias cosmol\'ogicas y la distribuci\'on de las componentes oscuras en el universo. 
Precisamente, las lentes gravitacionales son  una herramientas extremadamente \'util para medir la   distribuci\'on total del universo  y poder observar objetos muy  distantes. Esta t\'ecnica es importante  para la medici\'on precisa de masa de c\'umulos gal\'acticos y del estudio de  las propiedades f\'isicas de galaxias que est\'an en corrimientos al rojo intermedios entre la lente y el observador, \cite{bartelmann}.  Tambi\'en realiza mediciones directas del par\'ametro de Hubble, distribuciones de masa radiales en superc\'umulos y estimaci\'on de par\'ametros cosmol\'ogicos. 
El objetivo de este  art\'iculo es discutir  la estimaci\'on de la funci\'on de correlaci\'on de dos puntos en el cizallamiento gravitacional producido por la estructura a gran escala del universo (usualmente conocido como \emph{cosmic shear} $\gamma$),  \cite{Blanford}. 
Estad\'{\i}sticas de segundo orden provenientes del cizallamiento cosmol\'ogico como la apertura de masa $M_{\textit{ap}}(\vec{\theta})$,la varianza del shear  $\langle{\gamma^2(\vec{\theta})}\rangle$ junto con la posibilidad de estudiar no-gausianidad en el campo de densidad de materia son tambi\'en brevemente discutidas. Durante el art\'iculo, se muestran algunos c\'odigos muy \'utiles a la hora de realizar c\'alculos generados por efecto de lente gravitacional. Este art\'iculo esta dividido de la siguiente forma, en la secci\'on 2, se introduce el concepto de lente gravitacional, algunos efectos y definiciones de variables importantes para el estudio de las distorsiones en im\'agenes, en la seci\'on 3, se muestra las  caracteristicas m\'as relevantes en el efecto de lente gravitacional d\'ebil y como a partir de la forma de las im\'agenes se mide el shear gravitacional. En la secci\'on 4 se describe la funci\'on de correlaci\'on del shear y la apertura de masa el cual da cuenta de el shear tangencial, y a partir de estos resultados se muestra la forma en la cual se conduce a la estimaci\'on de par\'ametros cosmol\'ogicos, en particular de $\Omega_m$ y $\sigma_8$. Finalmente, las conclusiones son  mostradas en la secci\'on 5.     

\section{Teor\'ia de lente gravitacional}

Suponga que existe una fuente luminosa en una posici\'on $\mathbf{\eta}$, cuya luz emitida pasa a trav\'es de  una distribuci\'on de masa y con un par\'ametro de impacto $\mathbf{\xi}$ (ver figura \ref{fig1}). Por consideraciones geom\'etricas,  se puede obtener la deflexi\'on del rayo respecto al incidente debida al efecto del campo gravitacional de la lente,  \cite{fig1}.
Si comparamos el tiempo que recorre este rayo de luz deflectado respecto a uno sin deflexi\'on, se observa que el primero debe ser mayor debido a que este recorre una trayectoria mayor. La diferencia entre estos tiempos viene dada por
\begin{equation}
\tau(\mathbf{\xi})=\frac{1+z_l}{c}\frac{D_{l}D_{s}}{D_{ls}}\left[\frac{1}{2}|\mathbf{\xi}-\mathbf{\eta}|^2 -\phi(\mathbf{\xi}) \right],
\end{equation}
donde $z_l$ es el redshift de la lente (distribuci\'on de masa) y $D_{l}$, $D_{s}$, $D_{ls}$ son las distancias diametrales angulares del observador a la lente, observador a la fuente y de la lente a la fuente respectivamente. El potencial gravitacional se obtiene a trav\'es de la soluci\'on de la ecuaci\'on (\ref{poison})  
\begin{equation}
\phi(\xi)=\frac{1}{\pi}\int\frac{\Sigma(\mathbf{y})}{\Sigma_{cr}} \ln|\mathbf{\xi}-\mathbf{y}|d\mathbf{y}, 
\end{equation}
con $\Sigma_{cr}=\frac{c^2D_s}{4\pi G D_l D_{ls}}$.  De otro lado, observando la figura (\ref{fig1}), se puede derivar la relaci\'on entre la posici\'on de la fuente ($\beta$) y la posici\'on de las im\'agenes
 ($\theta=\frac{\xi}{D_l}$), el cual puede ser escrito como
\begin{equation}\label{ec3}
\beta=\theta-\frac{D_{ls}}{D_s}\alpha(\mathbf{\xi}).
\end{equation}
Esta \'ultima ecuaci\'on es conocida como la \textit{ecuaci\'on de lente}. El \'angulo de deflexi\'on total (de una lente delgada o distribuci\'on proyectada en el plano de la lente) puede ser encontrado al sumar los \'angulos deflectados generados por varias  masas puntuales que genera la distribuci\'on.  En el l\'imite continuo se encuentra que el \'angulo viene dado por
\begin{equation}
\alpha(\mathbf{\xi})=\frac{4\pi G}{c^2}\int \kappa(\mathbf{\xi^\prime})\frac{\mathbf{\xi}-\mathbf{\xi^\prime}}{|\mathbf{\xi}-\mathbf{\xi^\prime}|^2},
\end{equation}
con $\kappa=\frac{\Sigma(\mathbf{\xi})}{\Sigma{cr}}$. Esta ecuaci\'on es v\'alida bajo ciertos supuestos de campo gravitacional d\'ebil, distribuci\'on de materia estacionaria y velocidad de la materia mucho menor  que $c$. 
Adem\'as de producir una deflexi\'on, la lente tambi\'en distorsiona y amplifica la imagen de una forma descrita por el tensor de magnificaci\'on
\begin{equation}\label{ec5}
\mathcal{M}_{ij}= \frac{\partial \beta_i}{\partial \theta_j}=\begin{pmatrix}
  1-\kappa -\gamma_1& -\gamma_2 \\
  -\gamma_2 &  1-\kappa +\gamma_1
 \end{pmatrix},
\end{equation}
donde  se uso el hecho de que el \'angulo de deflexi\'on puede ser expresado respecto al potencial gravitacional como una ecuaci\'on tipo Poisson
\begin{equation}\label{poison}
\nabla^2_{2D} \phi(\theta)=2\kappa(\mathbf{\theta}),
\end{equation}
con $\alpha=\nabla_\theta \phi$ y donde
\begin{equation}
\gamma_1=\frac{1}{2}\left(\frac{\partial^2\phi}{\partial\theta_1^2} -\frac{\partial^2\phi}{\partial\theta_2^2}\right), \quad 
\gamma_2=\frac{1}{2}\frac{\partial^2\phi}{\partial\theta_1\theta_2}.
\end{equation}
\begin{figure}[h!]
\centering
\includegraphics[width=100mm]{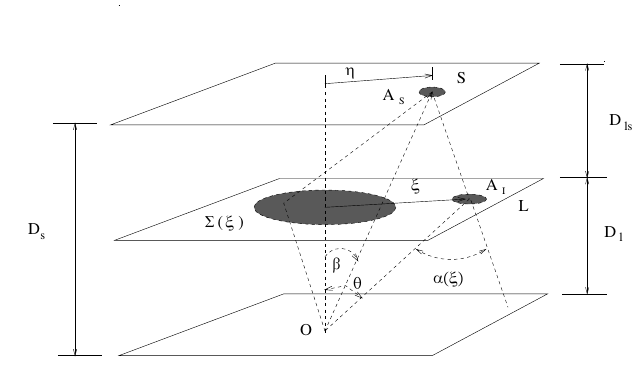}
\caption{Distancias  involucradas en un sistema de lente gravitacional. Figura tomada de \cite{fig1}.}
\label{fig1}
\end{figure}
La magnificaci\'on tambi\'en puede ser escrita de la siguiente forma
\[
\mu=\frac{1}{det \mathcal{M}}=\frac{1}{(1-\kappa)^2-\gamma^2}
\]
En  los lugares donde  $det\mathcal{M} = 0$ se tiene una magnificaci\'on infinita. 
Estas ubicaciones se denominan las curvas cr\'iticas. Para  
distribuciones  de  masa  esf\'ericamente sim\'etrica,   las curvas son c\'irculos,  para  puntos las  c\'austicas 
degeneran en  puntos y  para lentes el\'ipticas  
adem\'as de la cizalladura externa, las c\'austicas pueden constar de estrella y pliegues.
Usando programas computacionales como el gravlens, se puede determinar las caracteristicas que generan los modelos de masa propuestos, \cite{gravlens}. Por ejemplo si se modela la distribuci\'on de masa a trav\'es de una esfera singular isoterma, la presencia de imagenes viene dada como se observa en la figura \ref{fig2}.

\begin{figure}[h!]
\centering
\includegraphics[width=100mm]{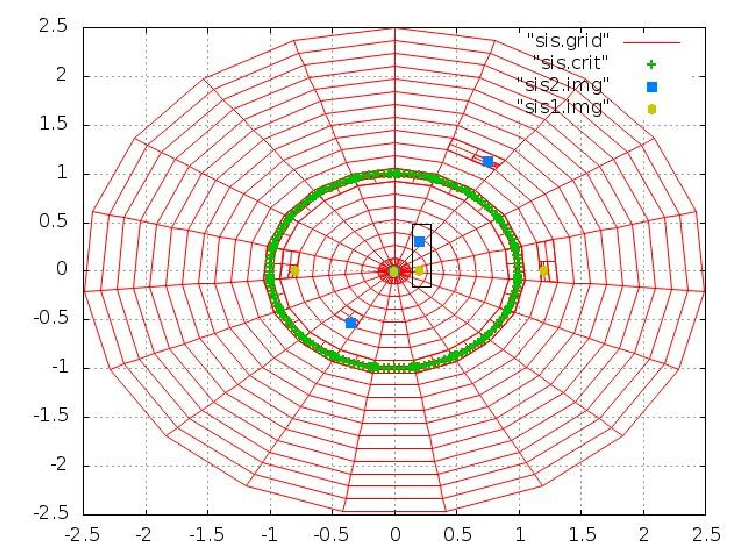}
\caption{Curvas cr\'iticas y c\'austicas para un modelo de masa esf\'erico singular isotermo. La malla esta en rojo, y la curvas cr\'iticas se muestra en verde, una de ellas esta en el origen de la malla. Existen dos fuentes (mostradas dentro del rectangulo negro) y las im\'agenes generadas debido al efecto de lente. }
\label{fig2}
\end{figure}
Para un modelo de masa el\'iptico singular isotermo, las im\'agenes generadas se muestran en la figura \ref{fig3}. 
\begin{figure}[h!]
\centering
\includegraphics[width=100mm]{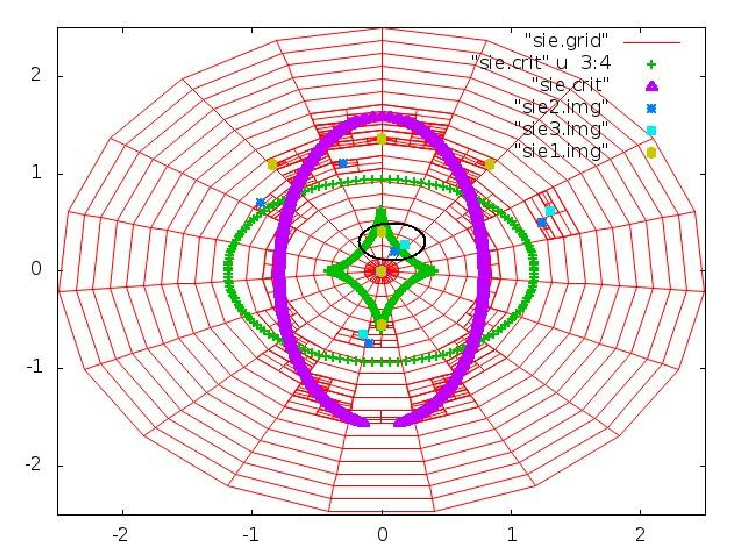}
\caption{Curvas cr\'iticas y c\'austicas para un modelo de masa el\'iptico singular isotermo. La malla esta en rojo,  la curvas cr\'iticas se muestra en magenta y las curvas c\'austicas se muestran en verde. Existen tres fuentes (mostradas dentro del \'ovalo negro) y las im\'agenes generadas debido al efecto de lente. }
\label{fig3}
\end{figure}
El campo gravitacional que genera el retardo temporal de las im\'agenes es mostrado en la figura \ref{fig4} y la correspondiente distribuci\'on de brillo en el plano de la lente es mostrado en la figura \ref{fig5}.
\begin{figure}[h!]
\centering
\includegraphics[width=70mm]{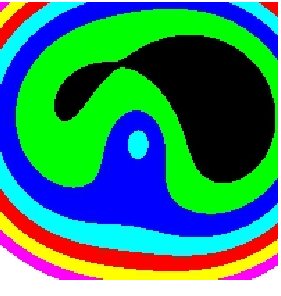}
\caption{Campo gravitacional generado en el modelo el\'iptico singular isot\'ermico. }
\label{fig4}
\end{figure}
\begin{figure}[h!]
\centering
\includegraphics[width=70mm]{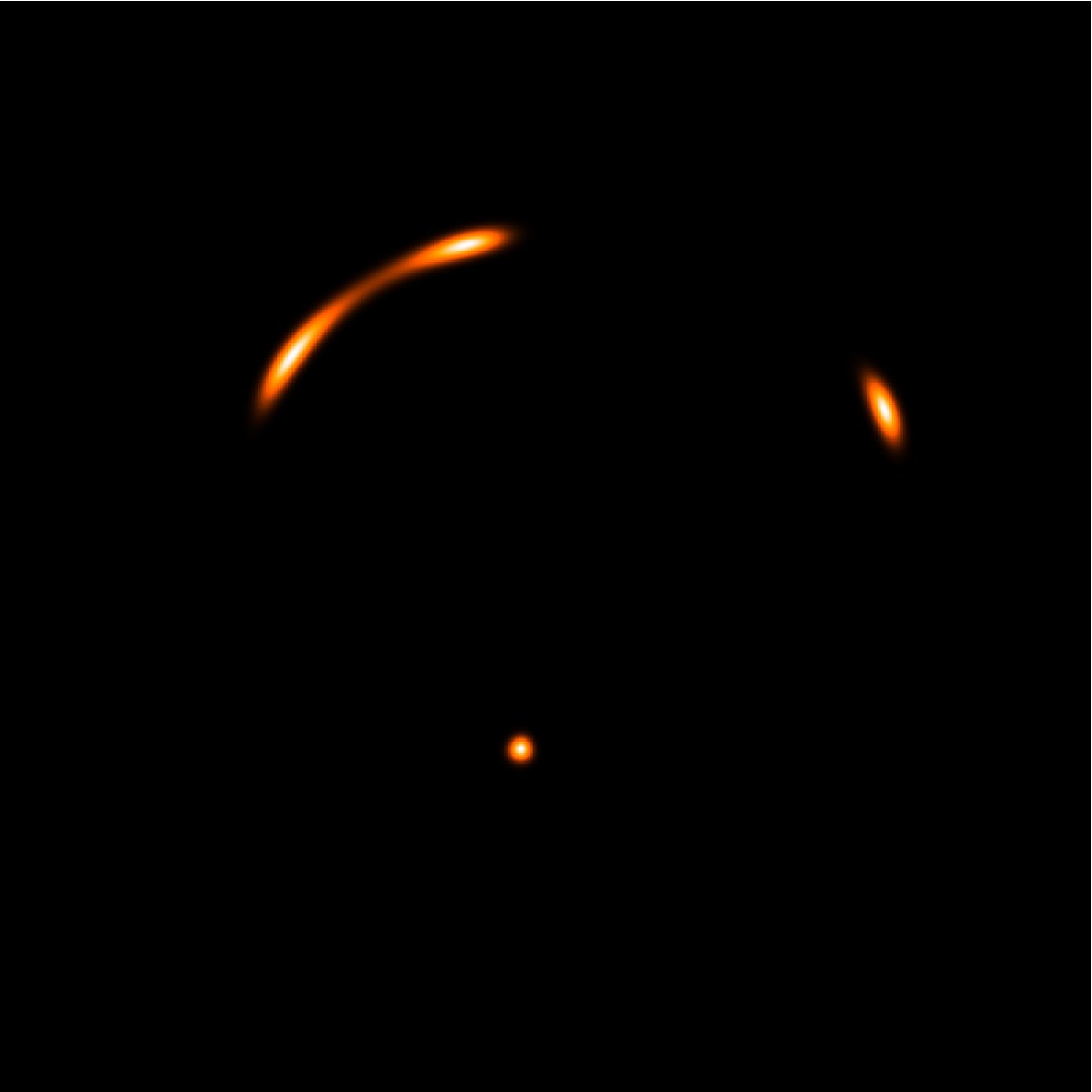}
\caption{Distribuci\'on de brillo en el plano de la lente para el modelo el\'iptico singular. }
\label{fig5}
\end{figure}
\section{Efecto de lente gravitacional en el r\'egimen d\'ebil}

Cuando hay cambios peque\~nos en la posici\'on de las fuentes, la distorsi\'on es d\'ebil (ver ecuaciones (\ref{ec3}) y (\ref{ec5})).
Uno de los principales efectos de la lente gravitacional d\'ebil son las distorsiones y la magnificaci\'on en el brillo sobre las im\'agenes de las galaxias.   
Estos efectos son generados via el cizallamiento gravitacional producido por la estructura a gran escala del universo (usualmente conocido como \emph{cosmic shear} $\gamma$). Dado esto, se puede inferir algunas caracteristicas debida al shear. Primero en el r\'egimen d\'ebil se tiene que $\kappa\ll 1$ y $\gamma \ll 1$, por lo tanto la magnificaci\'on y la distorsi\'on de las im\'agenes son pequeñas $\mathcal{M}\ll 1$. Segundo, la forma y la orientaci\'on (mediada por la elipticidad) de las im\'agenes son combinaciones tanto de la forma real de las galaxias mas la distorsi\'on debida al shear. Sin embargo, debido a que los efectos de distorsi\'on son peque\~nos, el problema se resume en el an\'alisis estad\'istico de todas las fuentes distorsionadas por efecto de lente gravitacional d\'ebil (ver figura \ref{fig1a}).
\begin{figure}[h!]
\centering
\includegraphics[width=80mm]{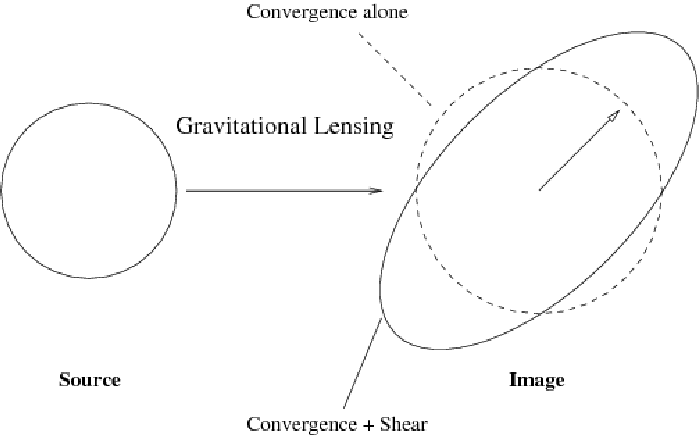}
\caption{Distorsi\'on de las im\'agenes debida al efecto de shear. }
\label{fig1a}
\end{figure}
De esta manera, se debe tener un estimador para las elipticidades de las galaxias y de all\'i poder determinar el shear en cada posici\'on. Se define la elipticidad de la forma
\begin{equation}
\epsilon_1+i\epsilon_2= \frac{Q_{11}-Q_{22}+2iQ_{12}}{Q_{11}+Q_{22}},
\end{equation}
donde $Q_{ij}$ es el brillo de luz el cual contiene la informaci\'on de la forma de las galaxias
\begin{equation}
Q_{ij}=\frac{\int d^2\theta q_I[I(\mathbf{\theta})](\mathbf{\theta}_i-\overline{\mathbf{\theta}}_i)(\mathbf{\theta}_j-\overline{\mathbf{\theta}}_j)}{\int d^2\theta q_I[I(\mathbf{\theta})]}.
\end{equation}
Bajo el tensor de magnificaci\'on, $Q_{ij}$ transforma de acuerdo a
\begin{equation}
Q^I=\mathcal{M}Q^S\mathcal{M}^T,
\end{equation}
con $I$ imagen y $S$ fuente.  Ahora debido a las propiedades de lente en el r\'egimen d\'ebil, la relaci\'on entre la elipticidad de fuente $\epsilon^S$ e im\'agen $\epsilon^I$ es aproximadamente dada por
\begin{equation}
\epsilon^I \approx \epsilon^S+ \gamma.
\end{equation}
Por lo tanto si se asume que la orientaci\'on de las galaxias es aleatoria (homogeneidad e isotrop\'ia del campo de elipticidades, indicando $<\epsilon^S>=0$) , el valor esperado de la elipticidad es
\begin{equation}\label{shearmedio}
<\epsilon^I> \approx \gamma.
\end{equation}
De esta forma se observa que al promediar sobre las elipticidades de las galaxias observadas, se obtiene el valor del shear.

\section{Determinaci\'on de par\'ametros cosmol\'ogicos}

Resultados arrojados por el WMAP o la misi\'on Planck, se ha encontrado que el universo esta formado en un $\sim 5\%$ de materia bari\'onica la cual tiene efectos importantes en  escalas  peque\~nas como los centros de galaxias y un $\sim 24\% $ de materia oscura el cual domina sobre estructuras  m\'as grandes.  
Debido a que la formaci\'on y evoluci\'on de estos sistemas depende en gran medida de lo que podamos saber de la materia oscura, debemos tener una manera de poder reconstruir esta estructura \textit{oscura}, \cite{Hoekstra}. La distribuci\'on de toda  la  materia   viene descrita por $\kappa$, pero esta variable puede ser medida a partir precisamente del shear. De esta forma, el efecto de lentes gravitacionales d\'ebiles nos dan una herramienta de poder estudiar la distribuci\'on de materia oscura en el universo. En el estudio de la lente gravitacional d\'ebil, $\kappa$ se calcula  usando la ecuaci\'on de inversi\'on de Kaiser-Squires
\begin{equation}\label{kapa1}
  \kappa(\mathbf{\theta})-\kappa_0=\frac{1}{\pi}\int_{\Re^{2}}d^2\mathbf{\theta}^\prime Re[D^*(\mathbf{\theta}-\mathbf{\theta}^\prime)\gamma(\mathbf{\theta}^\prime)],
\end{equation} 
 siendo $\kappa_0$ una constante y  donde el kernel esta definido por
\begin{equation}
D(\mathbf{\theta})=\frac{\theta_2^2-\theta_1^2-2i\theta_1\theta_2}{|\theta|^4}.
\end{equation}
Con esto,  podemos determinar el valor de $\kappa$ realizando la convoluci\'on del shear con el kernel. Sin embargo, el c\'alculo de $\kappa$ no es trivial, este supone algunas consideraciones y existen problemas de campo finito para realizar la integraci\'on de la ecuaci\'on (\ref{kapa1}), medici\'on de $\gamma$ entre otros.  
Ahora, si existe una fuente lejana que emite luz, esta radiaci\'on debe pasar por toda la distribuci\'on inhomog\'enea de masa que forma la estructura a gran escala del universo  por lo tanto, este rayo de luz sufre una distorsi\'on debida al efecto de lente  producido por esta inhomogeneidad (ver figura \ref{fig61}). 

\begin{figure}[h!]
\centering
\includegraphics[width=80mm]{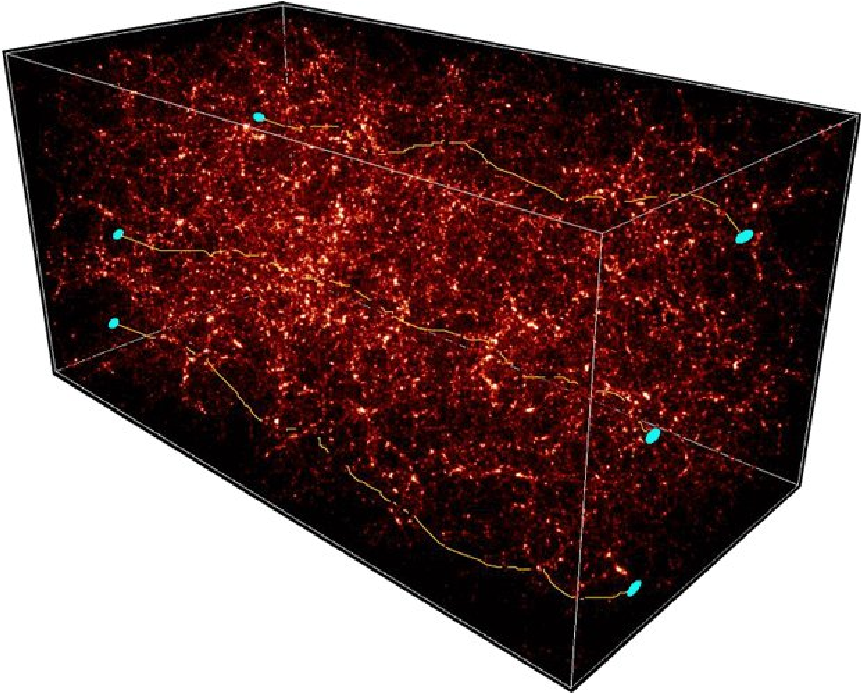}
\caption{Lente  gravitacional producido por estructura la gran escala c\'osmica. Copyright: Colombi $\&$ Mellier. }
\label{fig61}
\end{figure}
 Si se realiza un procedimiento estad\'istico de la elipticidad de las im\'agenes de las galaxias, se puede deducir todas las propiedades y caracteristicas de la estructura a gran escala y determinar la estimaci\'on de los par\'ametros de materia $\Omega_m$ y $\sigma_8$. En este caso el c\'alculo para $\kappa$ ecuaci\'on (\ref{kapa1}),  debe ser generalizado para que tenga en cuenta la distribuci\'on tridimensional de la materia.  
Para fuentes que tienen una distancia comovil $\chi$ y a un redshift $z$, el valor de $\kappa$ viene dado por

\begin{equation}
\kappa(\mathbf{\theta})=\frac{3H_0^2\Omega_m}{2c^2}\int_0^{\chi_{H}}  d\chi^\prime \overline{W}(\chi^\prime)f_k(\chi^\prime)\frac{\delta[f_k(\chi^\prime)\mathbf{\theta},\chi^\prime]}{a(\chi^\prime)},
\end{equation}
donde $a$ es el factor de escala y la distribuci\'on de fuentes
\begin{equation}
\overline{W}(\chi)=\int_\chi^{\chi_H}d\chi^\prime \frac{G(\chi^\prime)f_k(\chi^\prime-\chi)}{f_k(\chi^\prime)}, \quad G(\chi)d\chi=n_b(z)dz,
\end{equation}
  donde $\chi_H$ es la distancia de horizonte comovil,  $\delta$ la perturbaci\'on en la densidad en el espacio y  $f_k$ es la funci\'on de distribuci\'on de redshift 

\[
n_b(z)=\frac{\beta}{z_s\Gamma \left( \frac{1+\alpha}{\beta} \right) }\left(\frac{z}{z_s}\right)^\alpha\exp^{-\big(\frac{z}{z_s}\big)^\beta},
\]
con $z_s$ el redshift de la fuente $\Gamma$ la funci\'on Gamma y $\alpha$, $\beta$ par\'ametros libres.  
El observable de la estad\'istica de dos puntos es el espectro de potencias de la densidad de contraste (se puede usar el c\'odigo Athena 1.7 para c\'alculos de funciones de correlaci\'on a dos puntos, espectro de potencias y suavizado de funciones de segundo orden,  \cite{athena}). Para encontrar precisamente el espectro de potencias, se toma el  shear y se divide en una parte tangencial (t) y una componente transversal (c) 
\[
\gamma= \gamma_1+i\gamma_2, \quad \gamma_t=-Re(\gamma \exp^{-2i\psi}), \quad \gamma_c=-Im(\gamma \exp^{-2i\psi}),
\]
siendo $\psi$ el \'angulo polar y podemos  definir la correlaci\'on de dos puntos del shear como
\[
\xi_{\pm}\equiv <\gamma_t \gamma_t> \pm <\gamma_c \gamma_c> =\frac{1}{2\pi} \int_o^\infty dl l P_k(l)J_{0,4}(l\theta) ,
\]
siendo  $J_0$($J_4$) las funciones de Bessel correspondientes a la funci\'on de correlaci\'on  +(-),  \cite{Schneider2}. 
Una de las formas para medir el shear es con la estad\'istica de las elipticidades de las galaxias en una regi\'on determinada \textit{apertura}.  Para esto se considera una apertura circular de radio $\theta$ con un shear medio $\overline{\gamma}$ (ver ecuaci\'on (\ref{shearmedio})).
\begin{figure}[h!]
\centering
\includegraphics[width=80mm]{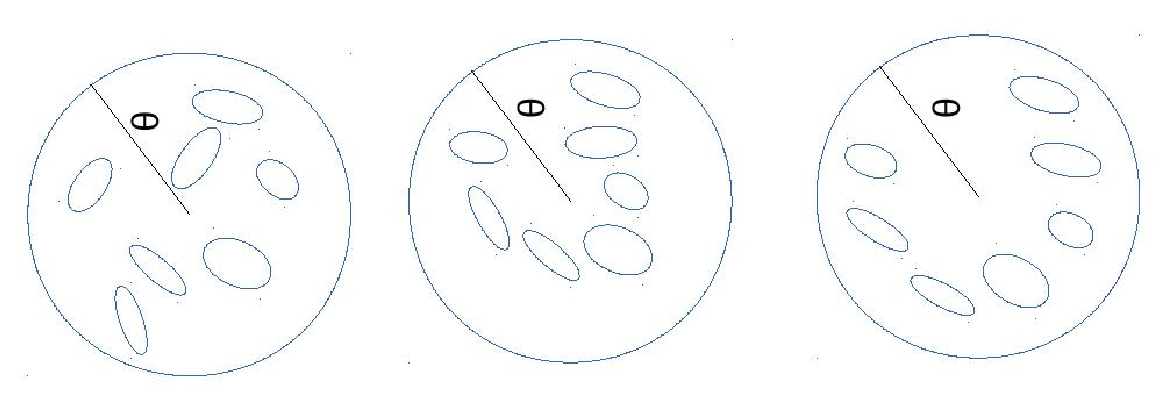}
\caption{Ensamble promedio de aperturas. }
\label{aper}
\end{figure}

 Si se realiza el ensamble promedio de estas aperturas (ver figura \ref{aper}), se puede definir la dispersi\'on del shear relacionado con el espectro de potencias viene dado por

\begin{equation}\label{shear1}
<|\gamma|^2>(\theta)=2\pi\int_0^\infty dl l P_k(l)\left[\frac{J_1(l\theta)}{\pi l \theta }\right]^2.
\end{equation}  
Tambi\'en se define la apertura de masa que es un estimador no cesgado  en el efecto de lente gravitacional
\begin{equation}\label{shear1}
<|M_{ap}|^2>(\theta)=\frac{1}{2 \pi}\int_0^\infty dl l P_k(l)576 \left[\frac{J_4(l\theta)}{ l \theta }\right]^2.
\end{equation}
 En una medida como $<M_{ap}^2>(\mathbf{\theta})$ se calcula la matriz de covarianza a diferentes escalas

\[
C_{ij}=<(d_i-\mu_i)^T(d_j-\mu_j)>,
\]

con $d_i$ la medida de $<M_{ap}^2>(\mathbf{\theta})$, $\mu_i$ el valor verdadero de $<M_{ap}^2>(\mathbf{\theta})$ (promedio sobre el ensamble a $\theta_i$).
Para un modelo dado $m_i$, se calcula
\[
\chi^2=(d_i-m_i)C^{-1}(d_i-m_i)^T,
\]
el cual es usado para calcular la verosimilitud de cada conjunto de par\'ametros. Desde $\chi^2$ se calcula los niveles de confianza.  Sin embargo, el valor que nos importa encontrar es el espectro de potencias de $\kappa$,  \cite{schneider}.  Para ello,  se define la funci\'on de correlaci\'on de dos puntos de $\kappa$  como
 \begin{equation}\label{ec1}
\xi_k(\mathbf{\theta}) \equiv \langle \kappa(\mathbf{\varphi})\kappa(\mathbf{\varphi}+\mathbf{\theta})\rangle,
\end{equation}
y la transformada de Fourier de la funci\'on de correlaci\'on de dos puntos es el espectro de potencias de $\kappa$
\begin{equation}
P_\kappa(l)=\frac{9H_0^4\Omega_m^2}{4c^4}\int_0^{\chi_H}d\chi^\prime\left(\frac{\overline{W}(\chi^\prime)}{a(\chi^\prime)} \right)^2P_\delta\left(\frac{l}{f_k(\chi^\prime)},\chi^\prime\right).
\end{equation}
El espectro de potencias de $\kappa$ no se puede medir directamente, sin embargo se encuentra que este espectro esta relacionado al espectro de potencias del shear,  ecuaci\'on (\ref{shear1}). 
Por \'ultimo, uno de aspectos mas importantes de las lentes gravitacionales d\'ebiles es que debido a que $P_\kappa$ es completamente dependiente del modelo, mediciones del shear pueden estimar par\'ametros cosmol\'ogicos
\begin{equation}
\sigma_8=\frac{1}{2\pi^2}\int k^2 dk P_{\delta}(k)\big[\frac{3J_1(k_pk)}{k_pk}\big],
\end{equation}
donde $\sigma_8$ es la fluctuaci\'on rms de la densidad a una escala de 8$Mpc$ (otra forma de normalizar el espectro usando conteo de galaxias viene descrito por \cite{martin}). La amplitud de cizallamiento cosmol\'ogico se puede ilustrar en el caso de un espectro de potencias escalada via una ley de potencias y una poblaci\'on de galaxias a cierto corrimiento al rojo. Si no se considera constante cosmol\'ogica, $\langle \kappa(\theta)^2\rangle$ se puede escribir como
\begin{equation}\label{ult}
\langle \kappa(\theta)^2\rangle \sim \sigma_8 \Omega_m^{0.75} z_s^{0.8}\left(\frac{\theta}{1^\prime}  \right)^{-(\frac{n+2}{2})},
\end{equation}
 con $n$ el indice del espectro de masa y $z_s$ el corrimiento al rojo de las fuentes.  De la ecuaci\'on (\ref{ult}) es en principio medible la combinaci\'on $\sigma_8$, $\Omega_m$. Esta degenerancia puede ser rota cuando ambos la varianza y la asimetr\'ia (skewness) de la convergencia son medidas,  \cite{tereno}.  
\begin{figure}[h!]
\centering
\includegraphics[width=100mm]{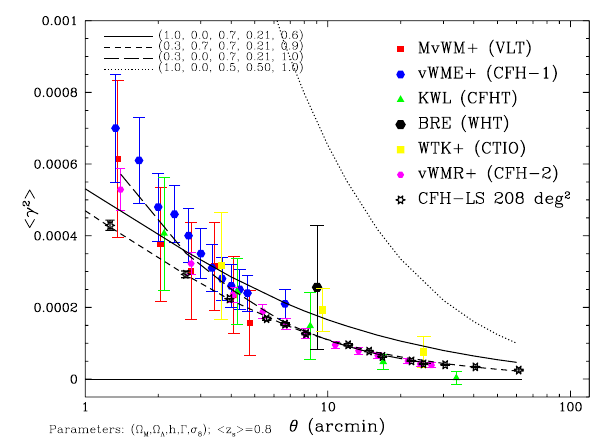}
\caption{Grafica de la varianza del shear en funci\'on de la escala angular para seis cat\'alogos de galaxias. Grafica tomada de \cite{tereno}. }
\label{ult1}
\end{figure}

La fuerte dependencia del shear con los par\'ametros cosmol\'ogicos ha sido ampliamente estudiada por varios autores, en particular \cite{tereno},  describe como es la varianza del shear en funci\'on de la escala angular para algunos modelos cosmol\'ogicos como se observa en la figura \ref{ult1}. 

\begin{figure}[h!]
\centering
\includegraphics[width=80mm]{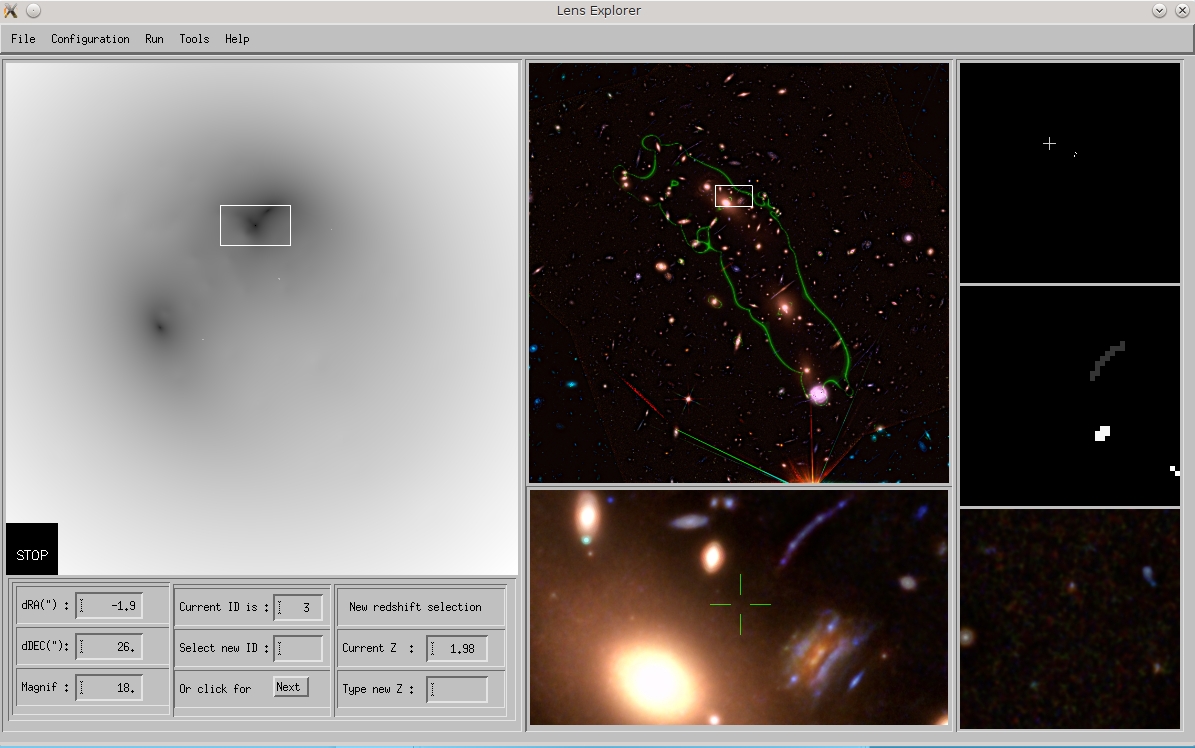}
\caption{Ventana de LensExplorer, donde se visualiza la formaci\'on de im\'agenes para el c\'umulo MACSJ0416.}
\label{lens1}
\end{figure}
\begin{figure}[h!]
\centering
\includegraphics[width=80mm]{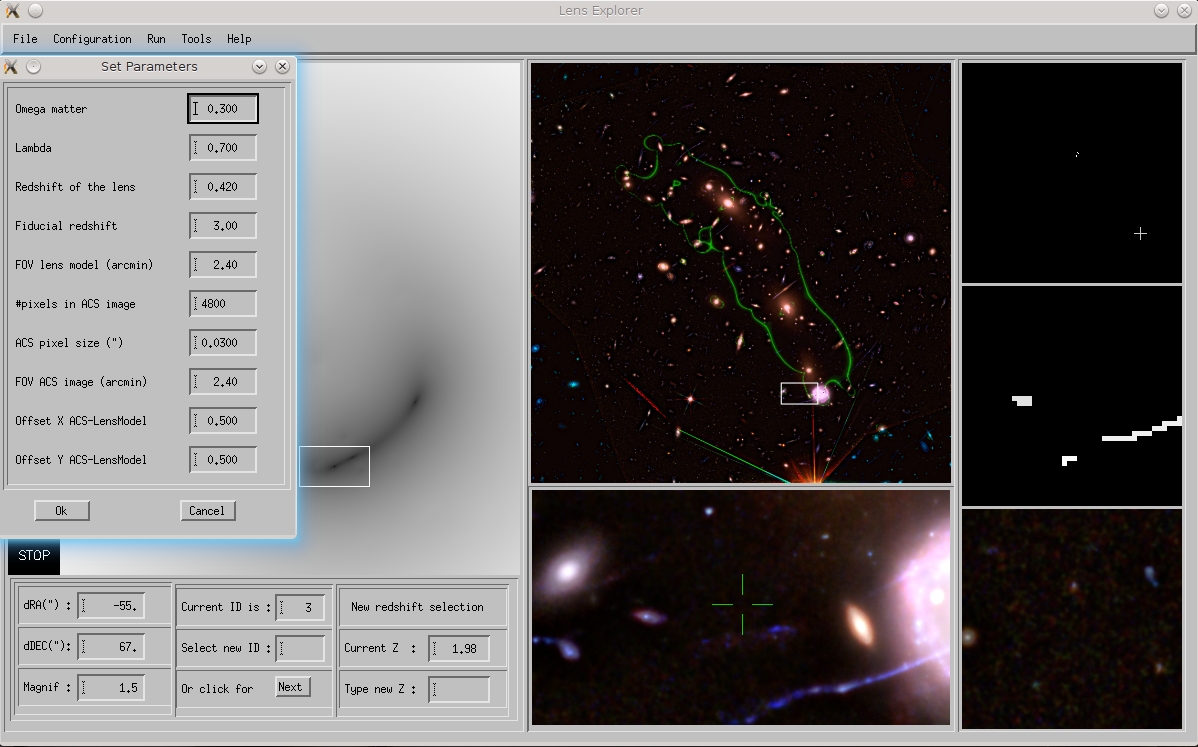}
\caption{Cambio de modelos, p\'arametros cosmologicos y corrimientos al rojo de la lente con LensExplorer.}
\label{lens2}
\end{figure}
 Otra forma de ver esta dependencia es usando algunos software para observar como varia el modelo de distribuci\'on de acuerdo a los par\'ametros cosm\'ologicos. Uno de ellos es LensExplorer descrito por \cite{lensexplorer}. En las figuras \ref{lens1} y \ref{lens2} se observa un ejemplo de como se puede visualizar la desviaci\'on por efecto de lente y predecir magnificaci\'on a diferentes corrimientos al rojo dependiendo de los par\'ametros cosmol\'ogicos. 

\section{Conclusiones}
El cizallamiento c\'osmico el cual distorsiona las im\'agenes de galaxias distantes por fuerzas de marea gravitacionales, 
ofrece una forma directa de estimar par\'ametros cosmol\'ogicos. 
La distribuci\'on de materia $\Omega_m$ y la normalizaci\'on del espectro de fluctuaciones en la densidad de materia $\sigma_8$ son especialmente trazados por el 
efecto de lente gravitacional en su r\'egimen d\'ebil. Para la estimaci\'on de dichos par\'ametros se emplea la correlaci\'on inducida por la gravedad en el campo de 
elipticidades gal\'actico. En este art\'iculo se ha mostrado que en el r\'egimen d\'ebil, el efecto de lente gravitacional ha proporcionado un camino directo para romper
el degeneramiento en la estimaci\'on de los par\'ametros cosmol\'ogicos, en especial $\Omega_m$ y $\sigma_8$. Tambi\'en este efecto reconstruye tridimensionalmente las 
fluctuaciones en densidad y en los observables, gracias a la topograf\'ia c\'osmica y a los corrimientos al rojo fotom\'etricos.  Por \'ultimo, 
medidas m\'as all\'a del cizallamiento c\'osmico, son datos prometedores para mejorar en un futuro la determinaci\'on de la forma de las galaxias y la estimaci\'on de 
las funciones de correlaci\'on.

\renewcommand{\refname}{Bibliograf\'ia}
\bibliographystyle{harvard}
\bibliography{Leonardo}

\end{document}